\documentclass{aa}
\usepackage{showyourwork}

\usepackage{txfonts}
\usepackage{graphicx}

\def\fdg{\hbox{$.\!\!^\circ$}}          

\begin{document}

\title{Improving \textit{INTEGRAL}/SPI data analysis of GRBs}

\titlerunning{Improving \textit{INTEGRAL}/SPI analysis for GRBs}

\author{
  Björn Biltzinger\inst{1}
  \and
  Jochen Greiner\inst{1}
  \and
  J. Michael Burgess\inst{1}
  \and
  Thomas Siegert\inst{1,2}
}

\institute{Max-Planck-Institut f\"ur extraterrestrische Physik, Giessenbachstra{\ss}e 1, D-85748 Garching, Germany
\and Institut f\"ur Theoretische Physik und Astrophysik, Universit\"at W\"urzburg,  Emil-Fischer-Str. 31, 97074 W\"urzburg, Germany }

\date{Received --; accepted --}

\abstract{
The spectrometer on the international gamma-ray astrophysics laboratory (\textit{INTEGRAL}/SPI) is a coded mask instrument observing since 2002 in the keV to MeV energy range, which covers the peak of the $\nu F\nu$ spectrum of most gamma-ray bursts (GRBs). Since its launch in 2008, the gamma-ray burst monitor (GBM) on board the \textit{Fermi} satellite has been the primary instrument for analysing GRBs in the energy range between $\approx$ 10 keV and $\approx$ 10 MeV. Here, we show that the spectrometer on board \textit{INTEGRAL}, named `SPI', which  covers a similar energy range, can give equivalently constraining results for some parameters if we use an advanced analysis method. Also, combining the data of both instruments reduces the allowed parameter space in spectral fits. The main advantage of SPI over GBM is the energy resolution of $\approx$ 0.2\% at 1.3 MeV compared to $\approx$ 10\% for GBM. Therefore, SPI is an ideal instrument for precisely measuring the curvature of the spectrum. This is important, as it has been shown in recent years that physical models rather than heuristic functions should be fit to GRB data to obtain better insights into their still unknown emission mechanism, and the curvature of the peak is unique to the different physical models.
To fit physical models to SPI GRB data and get the maximal amount of information from the data, we developed new open-source analysis software, {\tt PySPI}. We apply these new techniques to GRB 120711A in order to validate and showcase the capabilities of this software. We show that {\tt PySPI} improves the analysis of SPI GRB data compared to the \textit{INTEGRAL} off-line scientific analysis software ({\tt OSA}). In addition, we demonstrate that the GBM and the SPI data for this particular GRB can be fitted well with a physical synchrotron model. This demonstrates that SPI can play an important role in GRB spectral model fitting.
}


\maketitle

\section{Introduction}

Gamma-ray bursts (GRBs) are short transient bursts of $X-$ and $\gamma$-rays \citep{Klebesadel+1973} with a typical active time of between a few milliseconds and a few hundred seconds for the prompt phase.
After the prompt emission, there is a long-lasting afterglow phase when the ejecta interact with the interstellar medium \citep{MeszarosRees1997, afterglow}. GRBs are classified into two groups depending on the duration of the prompt emission \citep{shortlong}. Long GRBs (prompt phase $\gtrapprox$ 2s) have been associated with the collapse of massive stars \citep{SN1, Hjorth+2003, SN2}, whereas short GRBs are  caused by mergers of compact objects, such as neutron stars \citep{Eichler+1989, gw}. In both cases there is consensus that the progenitor events of GRBs produce jetted relativistic outflows, which consist of several shells with different velocities, resulting in internal shell collisions \citep{ReesMeszaros1994, Mochkovitch+1995}.

The true physical mechanism(s) responsible for the prompt emission of GRBs is a heavily debated topic \citep[for a review, see][]{KumarZhang2015}.
Two of the possible emission mechanisms are synchrotron \citep[e.g.][]{syn_shell, bosnjak_syn, synch} and photospheric emission \citep[e.g.][]{Goodman1986, photo_1, photo_2, photo_3}. In the past, GRB data were often fitted with empirical functions like the Band function \citep{band} in order to decipher the preferred physical model from the fit parameters. In recent years, it has been shown that this approach can be misleading and that it is preferable to fit physical models directly to the data \citep{Burgess-2014, physical_models, synch}. A prominent example of this is the so-called `line of death' for synchrotron emission \citep{line-of-death, line-of-death2}, which states that synchrotron radiation cannot be the universal emission mechanism if fits with a Band function give a low-energy power law slope larger than $-2/3$. \citet{synch} showed that this conclusion is flawed and that one can fit GRB spectra well with a physical synchrotron model even though the same spectra violate the line of death if fitted with a Band function. Another proposed proxy for physical emission processes in GRBs is the spectral curvature of empirical model fits to GRB data \citep{Yu-2015, Axelsson-2015}. This too has been shown to be an inaccurate indicator of the emission process \citep{Burgess2019}. Therefore, the use of physical models remains the best tool for deciphering the emission processes occurring in the relativistic outflows of GRBs.

In addition to the model used, the data also play an important role. Earlier analyses showed that the shape of the spectrum around the peak is a decisive feature. Therefore, we make use of the spectrometer (SPI) on the international gamma-ray astrophysics laboratory \citep[\textit{INTEGRAL};][]{integral} which provides 100 times higher spectral resolution in the 0.1--1 MeV band compared to the frequently used gamma-ray burst monitor (GBM) on board the \textit{Fermi} satellite.

The spectrometer SPI \citep{spi_main} is one of the instruments on board the \textit{INTEGRAL} satellite, which was launched in 2002. It is a coded mask instrument, with a fully coded field-of-view of 16 degrees, covering an energy range between 20 keV and 8 MeV. The unprecedented energy resolution allows the identification of fine spectral features, such as nuclear decay lines \citep{Roques-2003, Diehl-2006} and curvature in continuous spectra \citep{Jourdain-2020}.

The standard analysis of SPI GRB data is done within the \textit{INTEGRAL} Offline Scientific Analysis ({\tt OSA}) software \citep{osa}, including SPI-specific routines \citep{Diehl-2003}. A two-step method is applied, whereby first only the photo peak response is used to incorporate the mask pattern on the detectors, and then an energy redistribution correction response is applied. This method was used in several works that fitted empirical GRB models to the SPI data \citep{Malaguti-2003, Mereghetti-2003, Mereghetti-2003.2, Kienlin-2003, Kienlin-2003.2,Beckmann-2004, Moran-2005, Filliatre-2005.1, Filliatre-2005.2, McBreen-2006, Grebenev-2007, McGlynn-2008, Foley-2008, McGlynn-2009, Martin-Carrillo-2014}. The noteworthy exception of \citet{Bosnjak-2014} is discussed separately below in Sect. 2.5. Apart from data handling, all previous analysis of GRB data from INTEGRAL/SPI used heuristic functions like the Band function, but not a physical model.

Here, we developed the pure Python-based analysis tool {\tt PySPI} in order to provide an easy-to-use data-reduction and fitting tool while getting the maximal amount of information out of the data to progress to fitting physical models. This improves the analysis method in general, because we use the full response to forward model the expected spectra to the native SPI data space and use the correct likelihood to take the Poisson nature of the measurements into account. It also offers an  interface that is easy to install and use.
This work is structured as follows: In Sect. \ref{methods}, we summarise the different methods and concepts for SPI GRB data analysis, and provide a detailed description of the new analysis method within {\tt PySPI} in Sect. \ref{pyspi}. Section \ref{synch} introduces the physical synchrotron model we used in the fits and Sect. \ref{PPC} explains how we check whethre or not a fit is a good description of the data. We analyse the SPI and GBM data for GRB 120711A in Sect. \ref{analysis} and conclude our results in Sect. \ref{conclusion}.

\section{SPI GRB analysis methods}
\label{methods}
All instruments, especially gamma-ray instruments, suffer from energy dispersion and finite energy resolution. The information about this is encoded in the response matrix, which gives the probability that a photon with a certain energy and starting position on the sky will be detected in one of the electronic channels of the detector. These response matrices are usually not invertible, and therefore we have to forward fold the photon spectrum through the response matrix into the data space of the detected counts and compare them using the correct likelihood. This is a general statement that also applies to SPI. In the following subsection, we first cover some general concepts for SPI and then the standard analysis method within {\tt OSA,} as well as the method within our new analysis tool {\tt PySPI}. The main difference is that the method in {\tt PySPI} is a full forward-folding method, fulfilling the statement above, which results in maximising the information we can get from the data in the analysis.

\subsection{Response}
\label{response}

The response connects the photon spectrum to an expected detected count spectrum for a given source position. The response encapsulates all the information about the probability of a photon with a certain energy and origin to be detected in a certain electronic channel of the detector. This includes, for example, information about partial energy deposition of the photon in a crystal and the process that transforms the deposited energy into the electronic signal that is measured.
The response is split into two components, namely the image response functions (IRFs) and redistribution matrix files (RMFs). The IRFs contain information about the total effective area over which a photon can interact with the detector and the RMFs contain information about the probability in which electronic channel the photon will be measured.
For SPI, both these components were derived with extensive geometry and tracking (Geant) simulations \citep{spi_response}.
The SPI IRF files give the total effective area for three different interaction types:
 (1) photo peak events;
 (2) non-photo-peak events that first interact in the Ge crystal, and
 (3) non-photo-peak events that first interact in passive material.
These effective areas were calculated for 51 photon energies and on a 0\fdg5  grid out to 23\fdg5 from the on-axis direction. For each of the three interaction types, there is one RMF to define the shape of the redistributed spectra, assuming that this shape does not depend on the detector or the incident angles of the photons. The procedure used to construct the response includes several simplifications to keep the computational costs and storage space at manageable levels \citep{spi_response}. Re-simulating the response without these simplifications could improve the scientific output of SPI, but would be computationally very expensive, even today.

\subsection{Electronic noise}
\label{electronic}
\begin{figure}[ht]
    \begin{centering}
        \includegraphics{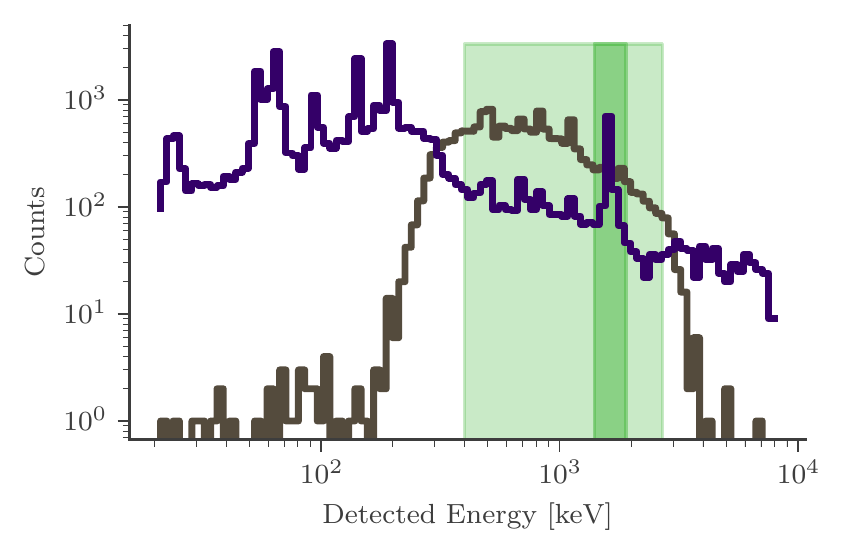}
        \caption{Count spectrum for detector 0 integrated over 1000 seconds in revolution 1189. The non-PSD event count spectrum is drawn in purple and the PSD events in grey. The light green shaded area marks the energy range within the PSD energy thresholds \citep{spi_electronic_noise}. The feature is clearly seen in the non-PSD events at 1.5 MeV (stronger green shaded area) and is not visible in the PSD events.}
        \label{fig:electronic}
    \end{centering}
\end{figure}

According to \citet{spi_electronic_noise}, there are spurious events in the SPI data, which are photons with small energy (<100 keV) that get detected at a higher energy due to saturation of the analog front-end electronics (AFEE) by previous high-energy deposition. It is also known that these spurious events do not show up in events that also have a detection in the pulse shape discriminator (PSD) electronics. The reason for this is that the PSD electronics have a low-energy threshold of $\approx$450 keV and a high-energy threshold of $\approx$2700 keV \citep[the exact values have been changed a few times during the mission; ][]{spi_electronic_noise}. Therefore, only events that deposit more than this low-energy threshold in the germanium crystal can trigger this electronic chain, which eliminates the $<$100 keV events that are detected at the wrong energies by the AFEE.  In Fig. \ref{fig:electronic}, the feature at 1.5 MeV is nicely visible in the non-PSD events but is missing in the PSD events. Even though this problem is most significant in the area around 1.5 MeV, it is also important at lower energies $>$400 keV. The electronic noise is not stable and depends on the signal strength \citep{spi_electronic_noise}, and therefore it cannot be included in the response and has to be treated differently.

\subsection{\textit{INTEGRAL} OSA software}
\label{OSA}

As we are introducing new analysis software in this work, we want to compare it to the existing analysis software to check whether or not the results are in agreement. We briefly summarise the standard analysis tools for GRB analysis, which are part of the {\tt OSA} software \citep{osa}.

Analysing GRB data from SPI with the {\tt OSA} tools is a multi-step process. The first steps involve selecting the science window containing the GRB, the time intervals for active time and background time, and the energy bins for the analysis. One can either find the location of the GRB with SPIROS, which is an iterative source-removal algorithm for SPI data \citep{Skinner-2003}, or set it manually if it is already known. The next steps encompass the spectral fitting, which is done in another two-step procedure.

The first step in the spectral fitting with {\tt OSA} is to use only the photo-peak response of the detectors. These responses depend on the source position, as this defines for example the absorption by the mask. With these responses and the measured data in the individual SPI detectors, taking into account the different background rates, a photon flux is fitted individually per energy channel. These pseudo photon fluxes are not the final result, as at this point, energy dispersion and detector energy resolution have not yet been accounted for.
The pseudo photon flux data points are fitted with a spectral photon flux model and a correction response. The correction response, that is available through {\tt OSA}, depends on the energy bins that are used and should account for energy dispersion. In this step, the energy channels are fitted simultaneously, as it is impossible to incorporate energy dispersion in a way that fits every energy channel individually. This fitting can be done in different spectral fitting software, such as {\tt XSPEC} \citep{xspec} or the Multi-Mission Maximum Likelihood framework \citep[3ML;][]{3ML}.

\subsection{{\tt PySPI}}
\label{pyspi}
To fit SPI GRB data, we developed a new Python package {\tt PySPI} \citep{joss}.
Below, we summarise the main points of GRB analysis within {\tt PySPI}.

\subsubsection{Background}

\begin{figure}
    \begin{centering}
        \includegraphics{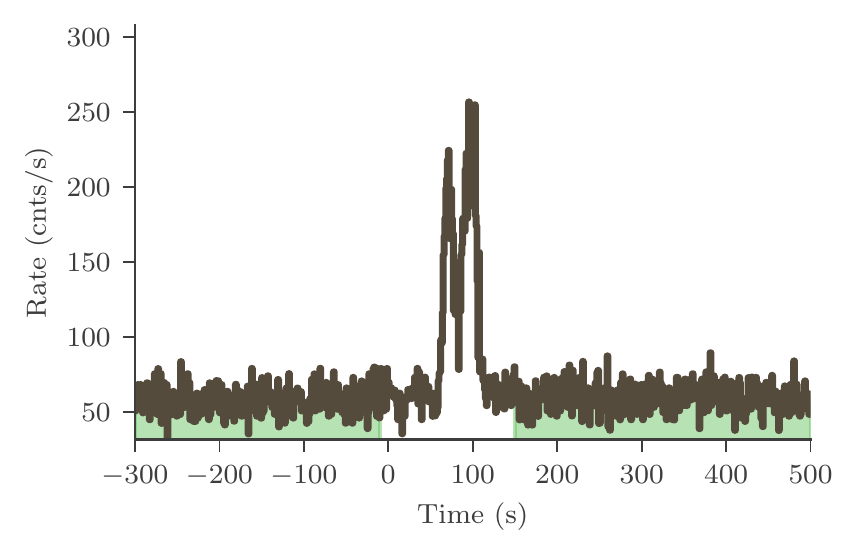}
        \caption{Light curve for SPI detector 0 for GRB 120711A. The transient source is clearly visible as well as the constant background for the time when the transient is not active. The time interval we use for the independent background observation is marked with the green shaded area.}
        \label{fig:lightcurve}
    \end{centering}
\end{figure}

GRBs are transient sources with a typical duration of up to a few tens of seconds. Therefore, we can use the time intervals during a single pointed observation (science window) ---when the transient source is not active--- as an independent temporal off-source observation (see Fig. \ref{fig:lightcurve}). This approach is similar to what is done with other instruments that use off-source observations to make independent background measurements.
The probability distribution for the background model rates per energy channel is the Poisson distribution:

\begin{equation}
        \mathcal{L}(B_{\mathrm{i}}, t_{\mathrm{b}}|b_{\mathrm{i}})=\frac{(t_{\mathrm{b}} b_{\mathrm{i}})^{B_{\mathrm{i}}}}{B_{\mathrm{i}}!}\cdot e^{-t_{\mathrm{b}} b_{\mathrm{i}}},
  \label{eq:poisson_bkg}
\end{equation}
where $b_{\mathrm{i}}$ are the background rates per energy channel, $B_{\mathrm{i}}$ are the detected counts in the off-source observation, and $t_{\mathrm{b}}$ is the exposure of the off-source observation.

\subsubsection{Likelihood}

With the defined background distribution, we can construct a likelihood, given by Eq. \ref{eq:likelihood_m_and_b}, that connects the source and background model with the Poisson process data of the background and active time interval via the response. Here, $\vec{\theta}$ summarises all the parameters of the source model, $D_{\mathrm{i}}$ are the measured counts in the selected active time interval of the transient source, $m_{\mathrm{i}}(\vec{\theta})$ are the predicted count rates from the model evaluated at the parameters $\vec{\theta,}$ and $t_{\mathrm{d}}$ is the exposure of the selected active time interval.

\begin{multline}
        \mathcal{L}(D_{\mathrm{i}}, B_{\mathrm{i}},t_{\mathrm{d}},t_{\mathrm{b}}|\vec{\theta}, b_{\mathrm{i}}) = \frac{(t_{\mathrm{d}}(m_{\mathrm{i}}(\vec{\theta})+ b_{\mathrm{i}}))^{D_{\mathrm{i}}}}{D_{\mathrm{i}}!}\cdot \\
  e^{-t_{\mathrm{d}}(m_{\mathrm{i}}(\vec{\theta})+b_{\mathrm{i}})}\frac{(t_{\mathrm{b}} b_{\mathrm{i}})^{B_{\mathrm{i}}}}{B_{\mathrm{i}}!} e^{-t_{\mathrm{b}} b_{\mathrm{i}}}.
  \label{eq:likelihood_m_and_b}
\end{multline}

\noindent
If there was a spectral model for the background, we would fit the background and the source model at the same time with the likelihood given in Eq. \ref{eq:likelihood_m_and_b}, but we do not have such a model for the SPI background. The background in SPI is dominated by the interaction of cosmic rays in the satellite material, which produces nuclear de-excitation line emission for example \citep{spi_bkg}. For this reason, the background consists of several hundred different nuclear lines \citep{spi_bkg}, which makes an accurate spectral model for the background impossible at present. However, we can marginalise over the parameters $b_{\mathrm{i}}$ using only the simple fact that they cannot be negative, which leaves us with Eq. \ref{eq:likelihood_marg}.

\begin{multline}
        \mathcal{L}(D_{\mathrm{i}}, B_{\mathrm{i}},t_{\mathrm{d}},t_{\mathrm{b}}|\vec{\theta}) = \int_{0}^{\infty}\textrm{db}_{\mathrm{i}}\frac{(t_{\mathrm{d}}(m_{\mathrm{i}}(\vec{\theta})+ b_{\mathrm{i}}))^{D_{\mathrm{i}}}}{D_{\mathrm{i}}!}\cdot\\
  e^{-t_{\mathrm{d}}(m_{\mathrm{i}}(\vec{\theta})+b_{\mathrm{i}})} \frac{(t_{\mathrm{b}} b_{\mathrm{i}})^{B_{\mathrm{i}}}}{B_{\mathrm{i}}!} e^{-t_{\mathrm{b}}b_{\mathrm{i}}}.
  \label{eq:likelihood_marg}
\end{multline}

\noindent
The marginalisation is equivalent to integrating out the background rate parameter assuming a uniform prior from zero to $\infty$. Because solving this integral is computationally expensive, we use a profile likelihood as an auxiliary but still statistically sound figure of merit. For the profile likelihood, we use the fact that the derivative of the likelihood at its maximum is zero ($\frac{\textrm{dL}}{\textrm{db}_i}=0$). This defines the background rates $b_\mathrm{{i, max}}$ (see Eq. \ref{eq:bmax}) that maximise the likelihood for a given model $m_{\mathrm{i}}(\vec{\theta})$ and observed quantities:
\begin{multline}
        b_{\mathrm{i,max}}(\vec{\theta})=\frac{1}{2(t_{\mathrm{b}}+t_{\mathrm{d}})}(B_{\mathrm{i}}+D_{\mathrm{i}}-m_{\mathrm{i}}(\vec{\theta})(t_{\mathrm{b}}+t_{\mathrm{d}})+\\
  \sqrt{(B_{\mathrm{i}}+D_{\mathrm{i}}+m_{\mathrm{i}}(\vec{\theta})(t_{\mathrm{b}}+t_{\mathrm{d}}))^{2}-4m_{\mathrm{i}}(\vec{\theta})D_{\mathrm{i}}(t_{\mathrm{d}}-t_{\mathrm{b}})}.
  \label{eq:bmax}
\end{multline}

\noindent
These values are then substituted into the likelihood Eq. \ref{eq:likelihood_m_and_b} to eliminate the $b_{\mathrm{i}}$ dependency. This gives the exact solution for the maximum of the likelihood; for likelihood values close to the maximum, the assumption is that most of the likelihood in the integrand in Eq. \ref{eq:likelihood_marg} is in a small area around $b_{\mathrm{i,max}}$.
This profile likelihood method has been used in many spectral analysis works \citep[e.g.][]{profile2, profile1} and is also available in {\tt XSPEC} \citep{xspec} as cstat and pgstat.

\subsubsection{Response}

In {\tt PySPI}, we use the official response files, as described in Sect. \ref{response}, interpolate them to the wanted energy bins and source positions, and construct one response matrix incorporating all the information about the effective area and energy redistribution. During this process, we take into account that the mask is not absorbing 100\% of the photons flying through it.

\subsubsection{Electronic noise}

In {\tt PySPI}, one can select the energy range in which only the PSD events should be used to avoid including spurious events (see Sec. \ref{electronic}). To account for the larger dead time of the PSD electronics, an effective area correction can be either fixed to 85\% \citep{spi_electronic_noise}, or treated as a free parameter of the model.

\subsubsection{General procedure}

Every Ge detector is treated as an independent detector unit. The workflow during a fit step is a forward folding method:
(1) sample model parameters, (2) calculate the model flux and responses individually for all detectors for the given source position, (3) fold the model with responses to get the predicted model counts in all detectors, (4) calculate the log-likelihood for all detectors, and (5) sum these log-likelihoods to get the total log-likelihood of the SPI data for the given model parameters.
The BALROG (BAyesian Lo-cation Reconstruction Of GRBs) algorithm for localising GRBs with \textit{Fermi}/GBM employs the same method \citep{balrog}, which shows that forward folding allows one to localise photon sources with a non-imaging instrument. It is important to realise that the instruments SPI and GBM are fundamentally very similar; they both consist of individual detectors that have different responses for a given source position. In GBM, this is due to the different pointing directions of the detectors, and in SPI it is due to the varying obscuration by the mask. Even though SPI is defined as a coded-mask instrument, it does not require special treatment. The information that the mask encodes into the data is automatically included by using the different responses for the detectors and a forward folding method. A similar approach was used by \citet{Tohuvavohu-2021} that introduces a new forward folding method to detect and localise faint GRBs with the coded mask instrument Swift/BAT.

\subsubsection{Software interface}

{\tt PySPI} constructs a plug-in for 3ML \citep{3ML}. This allows one to fit SPI data together with data from other instruments, such as \textit{Fermi}/GBM\footnote{See threeml.readthedocs.io}. {\tt PySPI} is open source software, and is publicly available on GitHub \citep{joss}, including documentation with examples\footnote{See pyspi.readthedocs.io}.

\subsection{Comparison of our method to Bo{\v{s}}njak et al. 2014}

\citet{Bosnjak-2014} performed a joint analysis of SPI and IBIS (Imager on Board the \textit{INTEGRAL} Satellite) data of GRBs and created a spectral catalogue of \textit{INTEGRAL} GRBs \citep{Ubertini-2003}. \citet{Bosnjak-2014} construct a response for every SPI detector and forward fold the photon model directly into the data space of every detector, which is similar to our approach in this work. Following their paper, we see two main differences compared to our method, namely response generation and treatment of the electronic noise, and one ill-defined issue (background treatment).

For the response generation, \citet{Bosnjak-2014} state that they calculate a response function taking into account the exposed fraction of each detector for the GRB position and that  the net
spectrum is zero for a completely shadowed detector. Therefore, we conclude that they used an idealistic mask with zero transparency at all energies. We on the other hand used the official response simulations that also include some transparency, especially at high energies. The second point is that the electronic noise is ignored completely in \citet{Bosnjak-2014}. \citet{spi_electronic_noise}  showed that the effect of the PSD events gets stronger for bright sources, and therefore it is important to account for this effect in the case of GRBs.

Concerning the background treatment, \citet{Bosnjak-2014} argue that background subtraction is better than using a model template for the background within {\tt OSA}, and cite  \citet{Cash-1979} for their likelihood. Therefore, one would conclude that they used a Poisson likelihood on the background-subtracted data, but they also state that they provide on-burst and background spectra separately to XSPEC, which implies that no background subtraction was done before the fit.  This was clarified with {\v{Z}}. Bo{\v{s}}njak in a private communication during the refereeing process: in XSPEC, background and burst(-source) spectra were used separately, and no background subtraction was done for the fits.

Unfortunately, the software used to generate the data files and response files in \citet{Bosnjak-2014} is not public and only the results for combined fits of SPI and IBIS data are shown. Therefore, we cannot directly compare our method and results with theirs. Also, \citet{Bosnjak-2014} only used data up to 1 MeV for SPI, whereas with PySPI we can use the data up to higher energies (up to 8 MeV), which is needed to constrain the peak energy for GRB 1200711A, which is at $\approx$ 1.4 MeV (see Sec. \ref{pyspi_gbm}).

\section{Physical synchrotron model}
\label{synch}
We use {\tt pynchrotron}\footnote{https://github.com/grburgess/pynchrotron} to calculate the spectrum from a physical spectral synchrotron model. This model was previously used to successfully fit many single-pulse GRBs observed with \textit{Fermi}/GBM in \citet{synch}, which also includes a detailed description of the model. Here, we summarise the main points.

The core of the model is the assumption of some generic acceleration mechanism that constantly accelerates electrons into a power-law spectrum $N(\gamma )\propto \gamma^{-p}$ between $\gamma_{\textrm{inj}}$ and $\gamma_{\textrm{max}}$. These electrons are cooled via the emission of synchrotron photons in a magnetic field. The resulting photon spectrum is the sum of the photon spectra at every time-step in the cooling process and is defined by five quantities: (1) the magnetic field strength ($B$), (2) the slope of the injected electron spectrum ($p$), (3) the lower boundary of the injected electron spectrum ($\gamma_{\mathrm{inj}}$), (4) the upper boundary of the injected electron spectrum ($\gamma_{\mathrm{max}}$), and (5) the characteristic Lorentz factor the electrons cool to ($\gamma_{\mathrm{cool}}$). There exists a strong degeneracy between $B$ and $\gamma_{\mathrm{inj}}$, as their combination sets the peak of the photon spectrum. Therefore, we fix $\gamma_{\mathrm{inj}}=10^{5}$ and only fit for $B$.

\section{Model checking}
\label{PPC}
How to decipher whether or not the fit of a model is a good description of the measured data is a very complicated and highly debated  topic in the statistics community. Measurements like reduced $\chi^{2}$ rely on many assumptions; for example, that all the probability distributions in the problem are described as normal distributions \citep{dosanddonts}. This is not the case here, as the measurement process is a Poisson process. We therefore decided to use posterior predictive checks (PPCs) and quantile-quantile (QQ) plots.

For the PPCs, one simulates new data from the full posterior of the fit and the measurement process (see Eq. \ref{eq:ppc}) and compares them to the observed data \citep{ppc}. QQ plots use the same simulation process, but instead of comparing the observed data of every energy channel individually to the simulated data, one compares the cumulative counts of the observation and the simulations over energy channels  \citep{QQ}. QQ plots are very sensitive to weak systematic deviations of the model from the data.

\begin{equation}
  \textrm{P}(y^{\textrm{sim}}|y^{\textrm{obs}}) = \int \textrm{P}(y^{\textrm{sim}}|\vec{\theta}) \textrm{P}(\vec{\theta}|y^{\textrm{obs}}) \mathrm{d}\vec{\theta}
  \label{eq:ppc}
.\end{equation}

\noindent
Here $y^{\textrm{obs}}$ are the observed data, $\textrm{P}(\vec{\theta}|y^{\textrm{obs}})$ is the posterior distribution of the model parameters $\vec{\theta}$ given the observed data, $\textrm{P}(y^{\textrm{sim}}|\vec{\theta})$ is the probability of new data given the parameters of the model, and $y^{\textrm{sim}}$ are the simulated data.

\section{Analysis}
\label{analysis}
\begin{figure}
  \begin{centering}
    \includegraphics[width=1\linewidth]{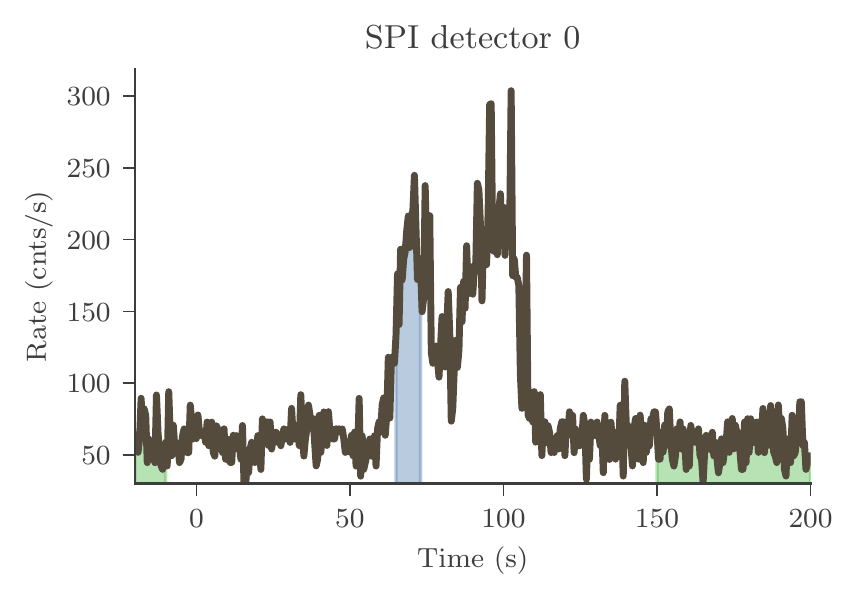}
    \includegraphics[width=1\linewidth]{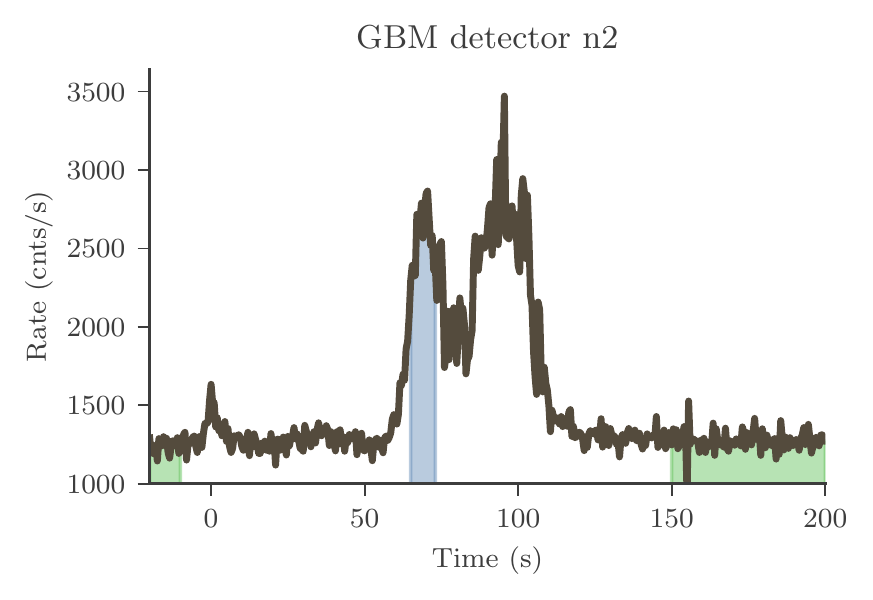}
    \caption{Light curve for GRB 120711A in one GBM and one SPI detector. The green shaded area marks part of the time intervals used for the independent background observation and the blue shaded area is the active time interval used in the fit.}
    \label{fig:time_selection}
  \end{centering}
\end{figure}

As an example, we analyse GRB 120711A, a bright, multi-pulse GRB with a precursor and a long emission period of $\approx$100 s. GRB 120711A was detected by SPI and GBM \citep{GCN_integral, GCN_gbm}.
We look at one time interval with a duration of 10 seconds around the first bright peak in the light curve (see Fig. \ref{fig:time_selection}). From the detection of the afterglow with the x-ray telescope (XRT) onboard the Swift satellite, we have an accurate position of the GRB (RA = 94\fdg68, Decl. = -71\fdg00 in J2000) \citep{GCN_swift}. The SPI data for GRB 120711A were previously analysed by \citet{Martin-Carrillo-2014}, but these authors only fitted the time-averaged spectrum over the $T_{90}$ emission (between 0 and 115 seconds after the trigger) with an empirical exponential cutoff power-law model, while focusing on the post-GRB emission properties.

First, we present our fit of the data with the empirical Band function, and show a comparison between the fit results using {\tt PySPI} versus {\tt OSA} (Sec. \ref{pyspi_osa}). We then check whether or not the results for SPI and GBM are in agreement (Sec. \ref{pyspi_gbm}). Subsequently, we present our fit of the SPI and GBM data simultaneously with a physical synchrotron model, which was done to check whether or not including SPI in the analysis can reduce the allowed parameter space (Sec. \ref{pyspi_gbm_joined}).

For all GBM and {\tt PySPI} fits, we added effective area correction parameters, allowing the total effective area of the individual detectors to vary with respect to each other in order to account for slightly different calibrations. To do this, we fixed the response of one of the detectors and fitted one parameter for every other detector. We constrain the effective area correction parameters to be between 0.7 and 1.3.
All the fits use {\tt 3ML} \citep{3ML} and the Bayesian sampling algorithm {\tt MultiNest} \citep{multinest} to create posterior distributions of the parameters.

\subsection{Comparing {\tt PySPI} and {\tt OSA}}
\label{pyspi_osa}
\begin{figure}
  \begin{centering}
    \includegraphics[width=1\linewidth]{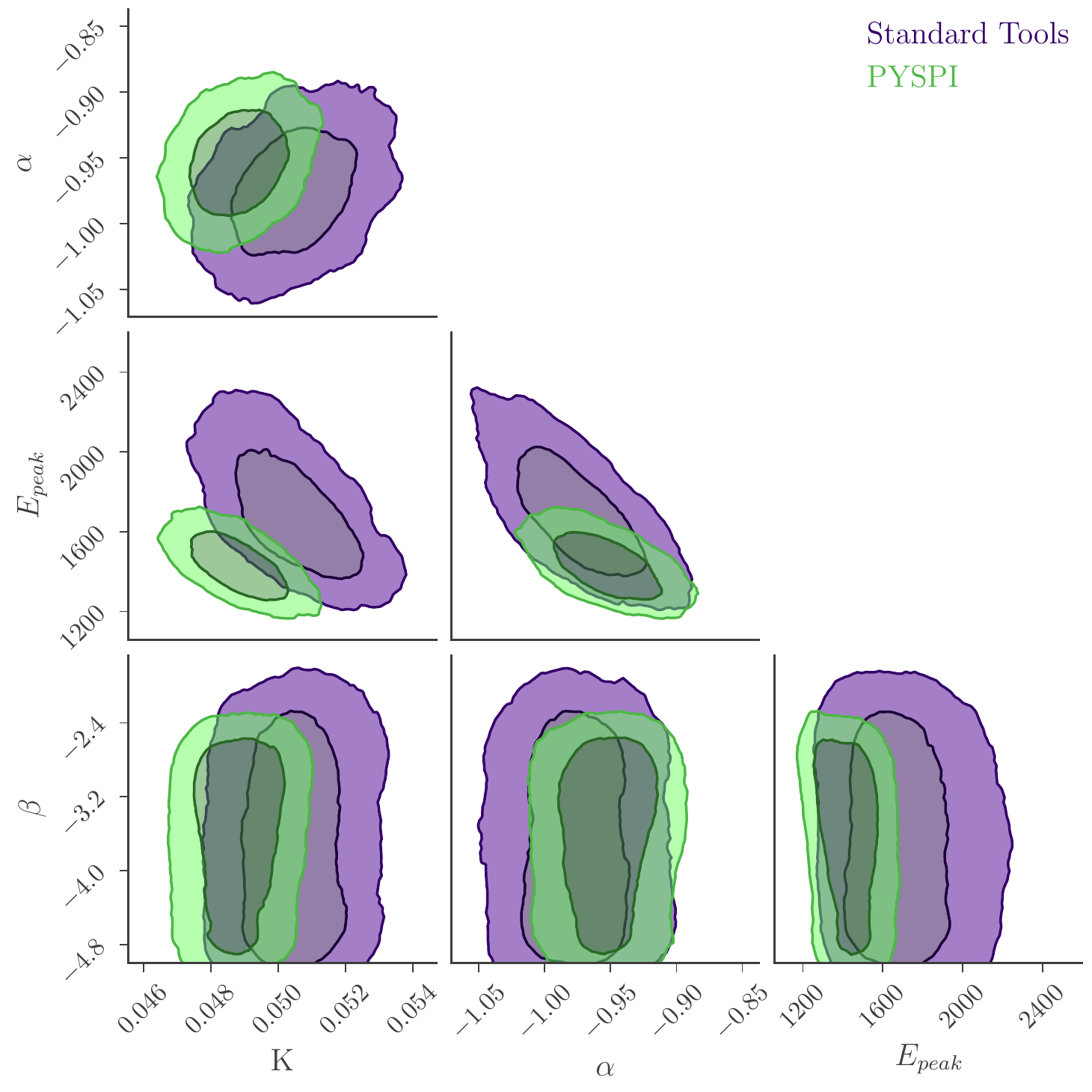}
    \caption{Corner plot showing the Band function fit to the SPI data with {\tt PySPI} and {\tt OSA} with the data for GRB120711A. The results are in agreement, but {\tt PySPI} constrains the parameters more precisely.}
    \label{fig:corner_osa_pyspi_band}
  \end{centering}
\end{figure}

\begin{figure*}
  \begin{centering}
    \includegraphics[width=0.45\linewidth]{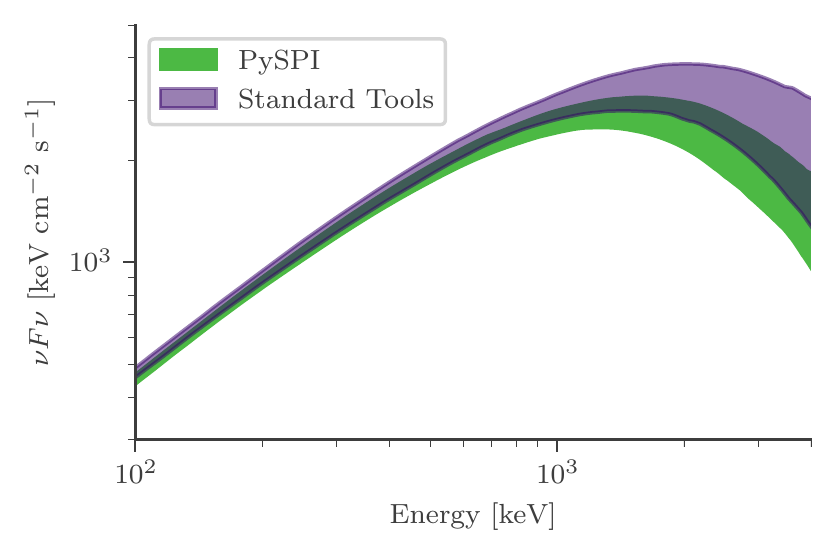}
    \includegraphics[width=0.45\linewidth]{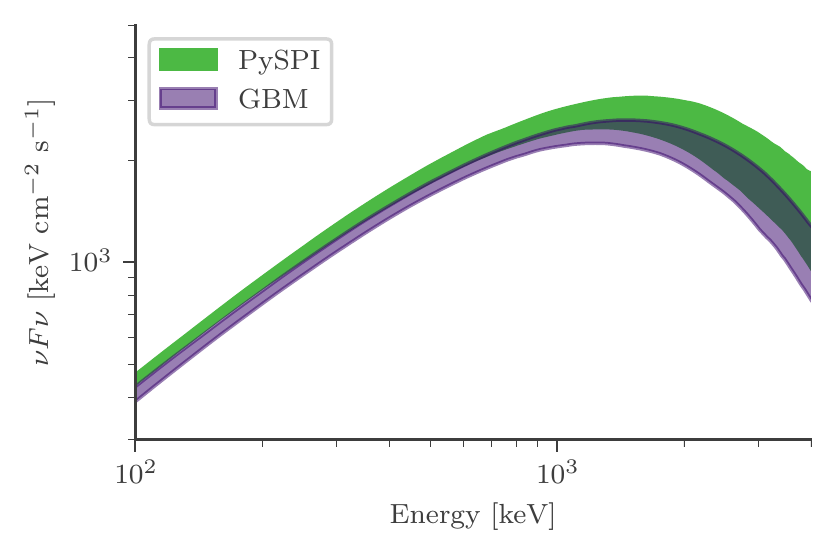}
    \caption{Model posterior plots (95\% confidence region) for the results with the Band function model and the data for GRB120711A. The left panel shows the results for the {\tt PySPI} fit compared to the fit using {\tt OSA} and the right panel shows the {\tt PySPI} fit compared to the GBM fit.}
    \label{fig:model_plot_band}
  \end{centering}
\end{figure*}

We fit the SPI data of an identical time interval with a Band function within {\tt PySPI} and {\tt OSA}.
Figure \ref{fig:corner_osa_pyspi_band} shows the resulting posterior distributions of the fits. The results for the spectral shape from the two analysis techniques agree within their uncertainty regions, but {\tt PySPI} can constrain the model more precisely (see Fig. \ref{fig:model_plot_band}).

\begin{figure}
  \begin{centering}
    \includegraphics[width=1\linewidth]{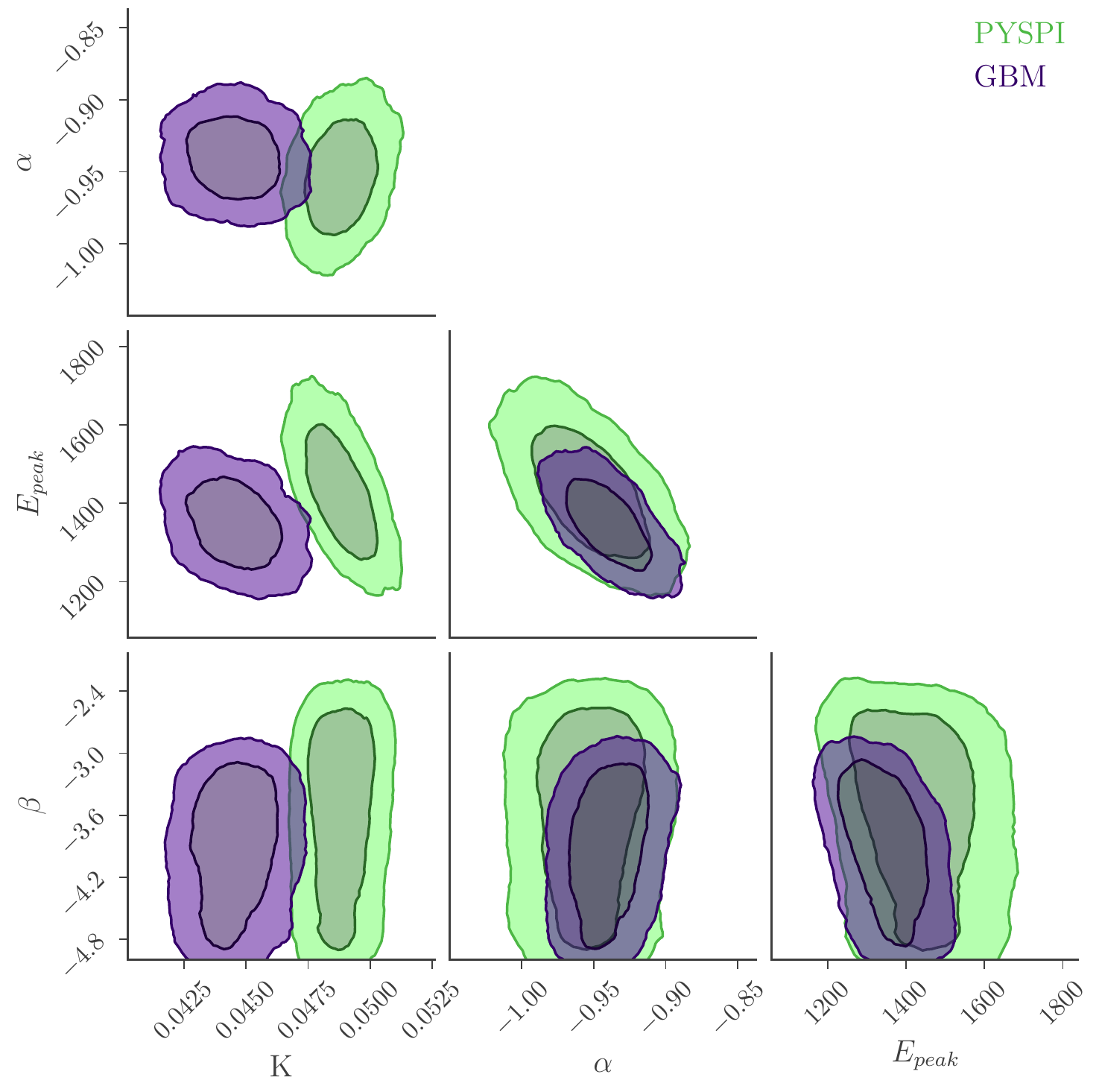}
    \caption{Corner plot for the SPI fit with {\tt PySPI} and the GBM fit to the data of GRB120711A. In this fit, we used a Band function as a spectral model. The spectral shapes for the SPI and GBM fits coincide within their uncertainty regions and the normalisation is off by $\approx$10\%.}
    \label{fig:corner_gbm_pyspi_band}
  \end{centering}
\end{figure}
\begin{figure}
  \begin{centering}
    \includegraphics[width=1\linewidth]{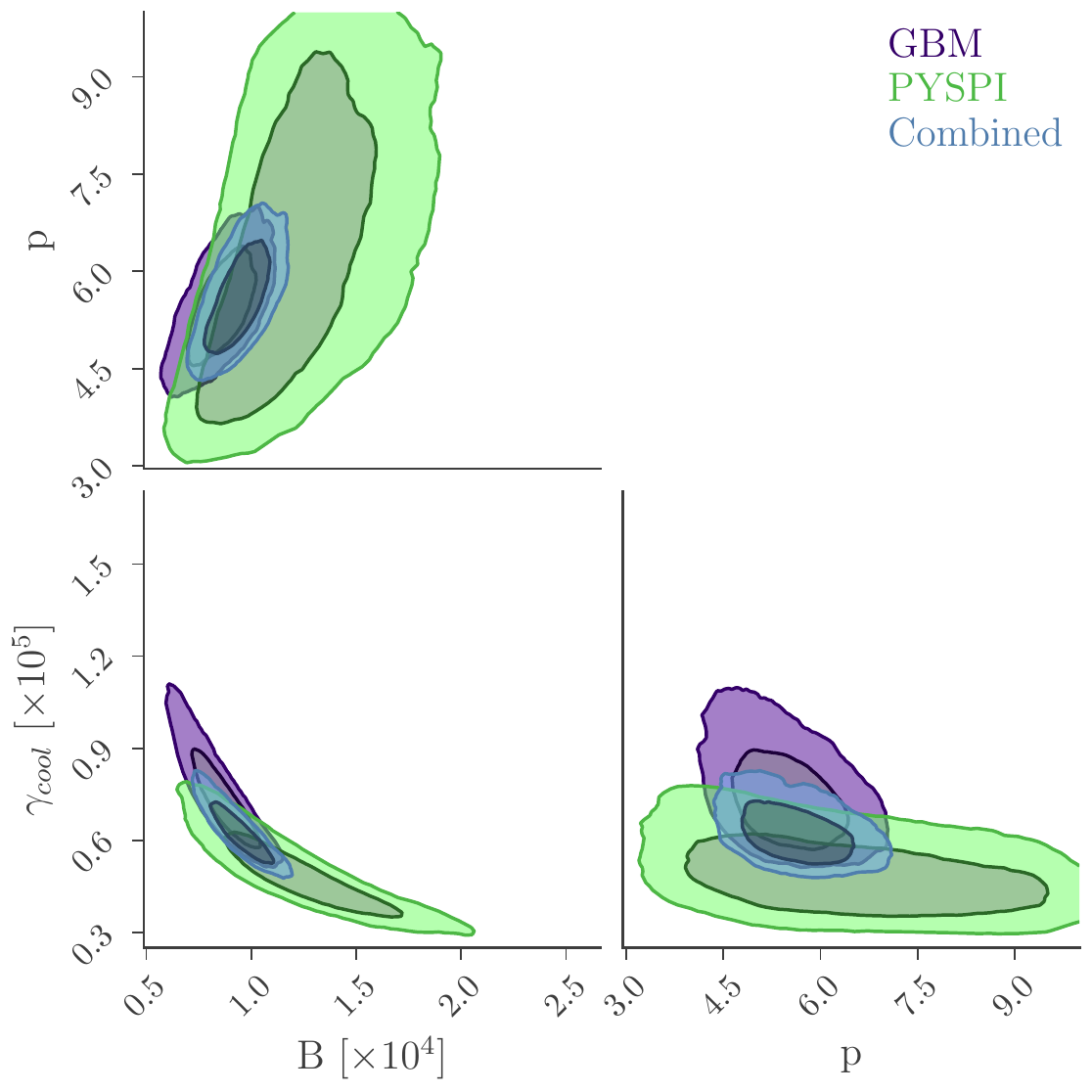}
    \caption{Corner plot for the SPI fit with {\tt PySPI} and the GBM fit to the data of GRB120711A. In this fit we used a physical synchrotron model (see Sec. \ref{synch}) as a spectral model. The results from SPI and GBM agree within their uncertainty region and the combined fit provides better constraints on the parameter than the individual ones.}
    \label{fig:corner_gbm_pyspi_joined_syn}
  \end{centering}
\end{figure}

\subsection{Comparing SPI and GBM}
\label{pyspi_gbm}
We then analysed SPI data with {\tt PySPI} and the GBM data with the GRB analysis within {\tt 3ML}, again with a Band function. Figure \ref{fig:corner_gbm_pyspi_band} shows the results for these fits; we can see that the results for the spectral shape from SPI and GBM are in agreement. It also shows that the effective area calibrations for SPI and GBM are well aligned, with a difference of  only $\approx$10\%
 ($K$ parameter), which is well within the
uncertainties as specified in the corresponding instrument publications.

\subsection{Joint fit of SPI and GBM}
\label{pyspi_gbm_joined}
We performed a joint fit of SPI and GBM data with a physical synchrotron model (see Sec. \ref{synch}). Figure \ref{fig:corner_gbm_pyspi_joined_syn} shows the corner plots for the individual and combined fits. The posterior distribution of the parameters from the GBM and the SPI fit agree within their uncertainty regions and the combined fit reduces the allowed parameter space. The physical synchrotron model was able to fit the data of GBM and SPI well, which is shown with PPC and QQ plots in Appendix \ref{appendix}.

\begin{figure*}
  \begin{centering}
    \includegraphics[width=0.45\linewidth]{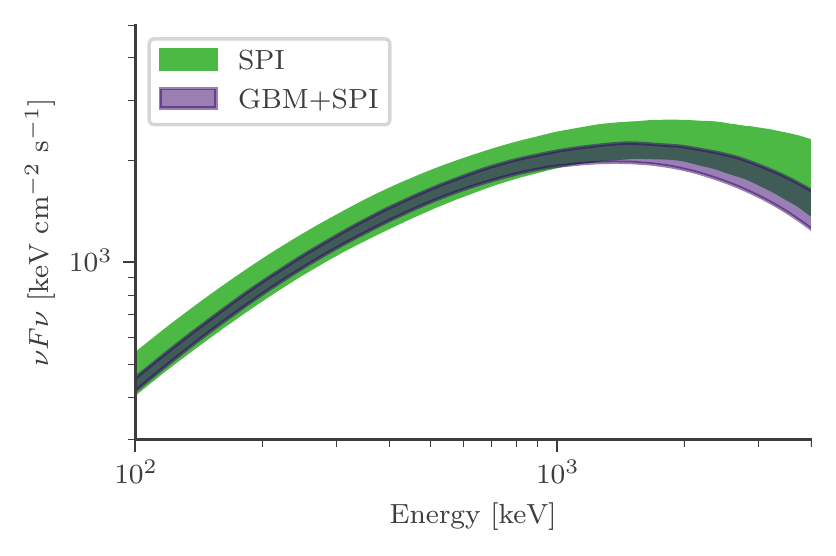}
    \includegraphics[width=0.45\linewidth]{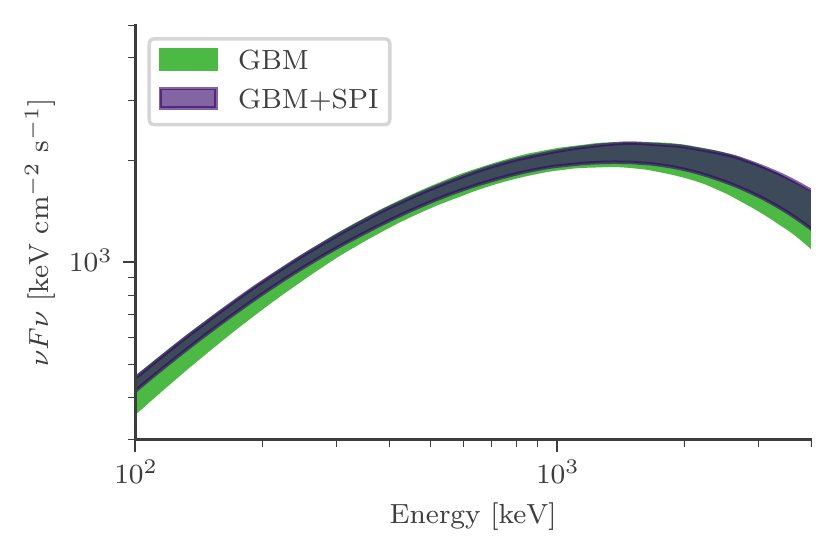}
    \caption{Model posterior plots (95\% confidence region) for the results with the physical synchrotron model and the data for GRB120711A. The left panel shows the results for the SPI fit compared to the combined fit and the right panel shows the GBM fit compared to the combined fit. The combined fit reduces the allowed model space.}
    \label{fig:model_plot_syn}
  \end{centering}
\end{figure*}
\begin{figure*}
  \begin{centering}
    \includegraphics[width=1\linewidth]{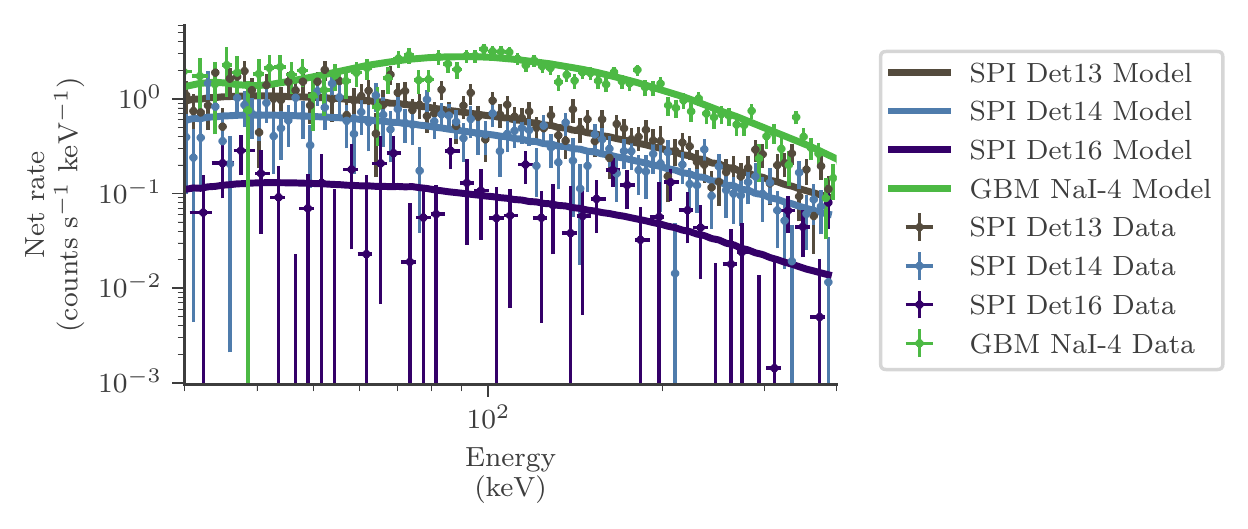}
    \includegraphics[width=1\linewidth]{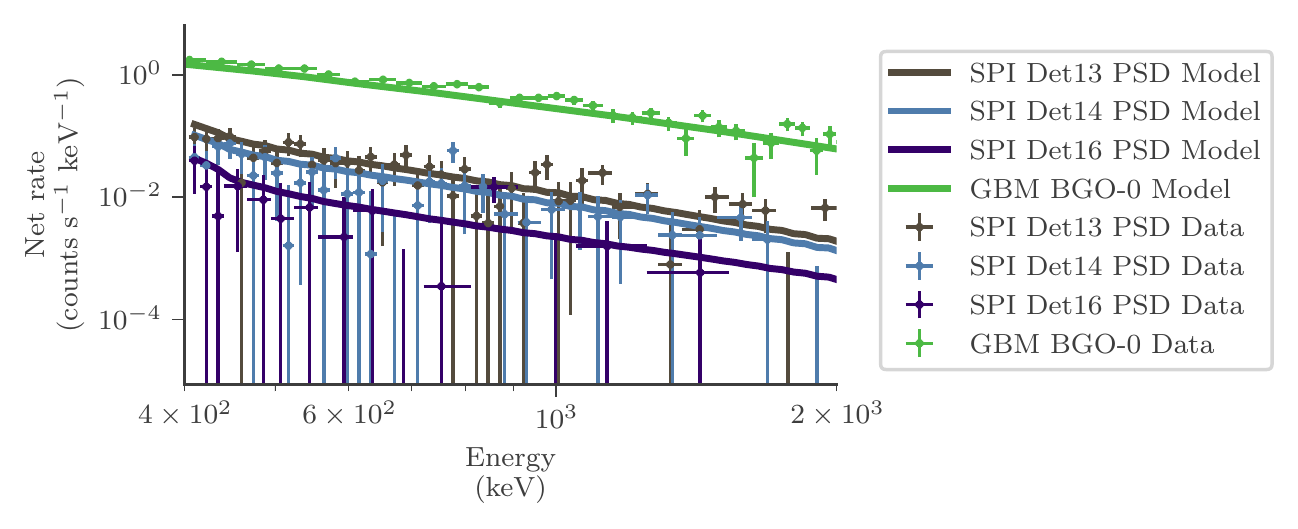}
    \caption{Data (for GRB120711A) and best-fit model in count space for three of the SPI detectors and one of the NaI (BGO) detectors of GBM  (upper  and lower plot, respectively). The spectral model for this fit was a physical synchrotron model.}
    \label{fig:dataplot}
  \end{centering}
\end{figure*}

\section{Conclusion}
\label{conclusion}
We present a new analysis method for GRBs detected by \textit{INTEGRAL}/SPI and the corresponding software package that uses this method. The {\tt PySPI} software uses a proper forward-folding technique for each detector (see Sec. \ref{pyspi}). To show that this improves the analysis of GRBs detected by SPI, we analysed GRB 120711A using the data from \textit{INTEGRAL}/SPI and \textit{Fermi}/GBM. The results for SPI with {\tt PySPI} are in agreement with those from the \textit{INTEGRAL} {\tt OSA} software, but provide better constraints. Also, the resulting spectral shape for the fits with GBM and SPI data are in agreement and the relative calibration difference between SPI and GBM ---which we conclude from the fits--- is $\approx$ 10\% or less. We show that the GBM as well as the SPI data for GRB 120711A can be fitted well with a physical synchrotron model and that combining the data in a simultaneous fit improves the parameter constraints. Consequently, SPI can play an important role in physical GRB model checking.
The next steps will entail an analysis of all GRBs detected by SPI with different physical models and {\tt PySPI} to check whether or not some of these models can be rejected with the SPI data. This will be covered in a future paper.

\begin{acknowledgement}
We thank {\v{Z}}. Bo{\v{s}}njak for quickly responding to our inquery about the details of data handling in \citet{Bosnjak-2014}. BB acknowledges support from the German Aerospace Center (Deutsches Zentrum f\"ur Luft- und Raumfahrt, DLR) under FKZ 50 0R 1913. TS acknowledges support by the Bundesministerium f\"ur Wirtschaft und Energie via the Deutsches Zentrum f\"ur Luft- und Raumfahrt (DLR) under contract number 50 OX 2201.
This work made use of the following software packages:\\
{\tt 3ML} \citep{3ML}, {\tt Astromodels} \citep{astromodels}, {\tt Matplotlib} \citep{matplotlib}, {\tt ChainConsumer} \citep{chainconsumer, chainconsumer2}, {\tt OSA} \citep{osa}, {\tt Numpy} \citep{numpy}, {\tt MultiNest} \citep{multinest, multinest1, multinest3}
\end{acknowledgement}

\bibliographystyle{aa}
\bibliography{bib}

\begin{appendix}
  \section{Localising GRB 120711A}

\begin{figure}[ht]
  \begin{centering}
    \includegraphics[width=0.7\linewidth]{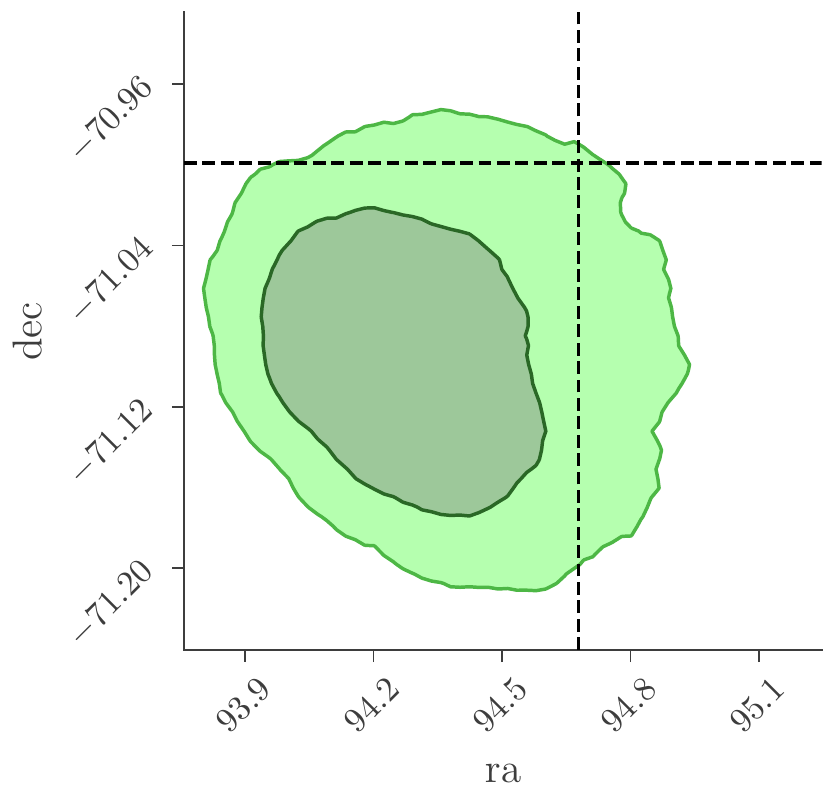}
    \caption{Localisation of GRB120711A with PySPI. Contours show the statistical only 1- and 2-sigma credible regions. We estimate an additional systematic error of $\approx$ 0.5 degrees because that is the resolution of the grid points in the response simulation. The dashed lines mark the position observed by Swift/XRT \citep{GCN_swift}.}
    \label{fig:localization}
  \end{centering}
\end{figure}

  It is also possible to localise GRBs with PySPI. To localise, we vary the spectrum parameters and the location of the source at the same time in one fit. For this, {\tt PySPI} has a fast response generator that uses the base response simulation grid to generate the response for any given source position quickly. Figure \ref{fig:localization} shows the localisation we get if we fit the location and use a Band function as spectral model. We also marked the position of the GRB measured by \textit{Swift}/XRT, which shows that the result with the SPI data is in agreement with the \textit{Swift}/XRT observation \citep{GCN_swift}. For this fit, we only used the energies up to 600 keV to avoid the PSD energy region.

\onecolumn
  \section{Model checking plots}
  \label{appendix}
  In this section we show a selection of the PPC- and QQ-plots for the simultaneous synchrotron fit to the SPI and GBM data for GRB120711A.

  \begin{figure*}[ht]
    \begin{centering}
      \includegraphics[width=0.45\linewidth]{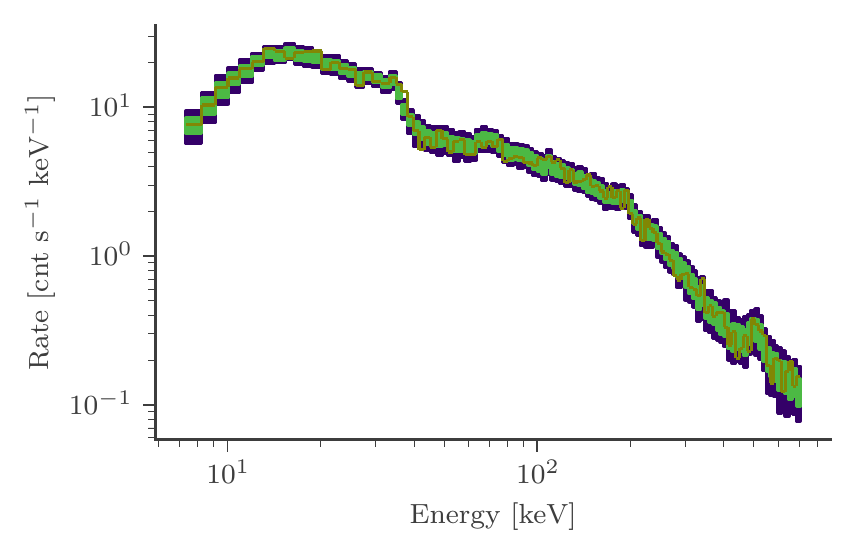}
      \includegraphics[width=0.45\linewidth]{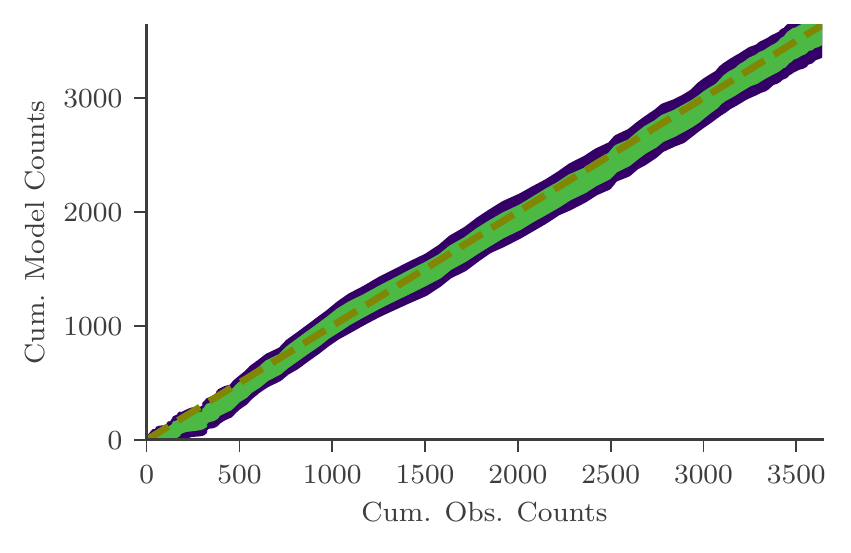}
      \caption{PPC (left) and QQ (right) plot for the GBM detector n0. The green and purple shaded areas are the one- and two-sigma contours. In the PPC plots, the dark yellow curve is the detected count spectrum, and in the QQ plots it shows the expected 1:1 relation between the cumulative model and observed counts.}
      \label{fig:n0_model_check}
    \end{centering}
  \end{figure*}
  \begin{figure*}[ht]
    \begin{centering}
      \includegraphics[width=0.45\linewidth]{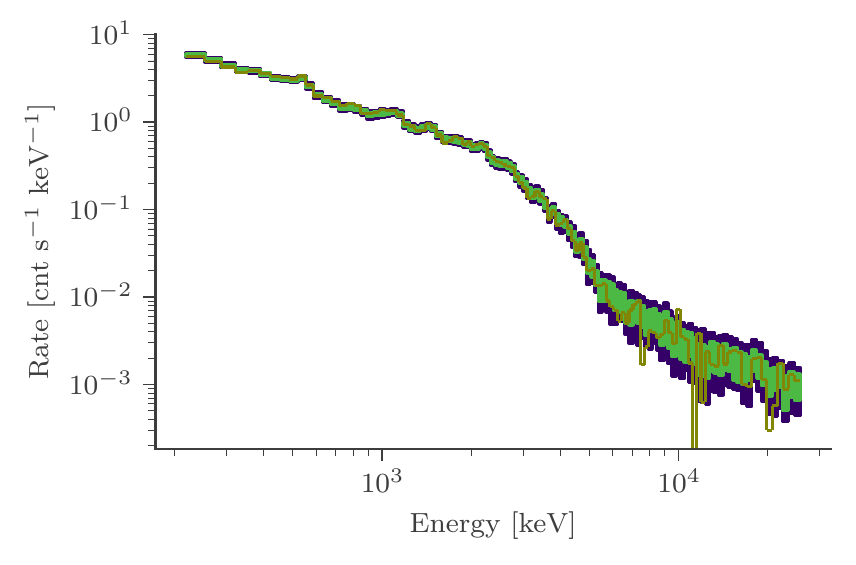}
      \includegraphics[width=0.45\linewidth]{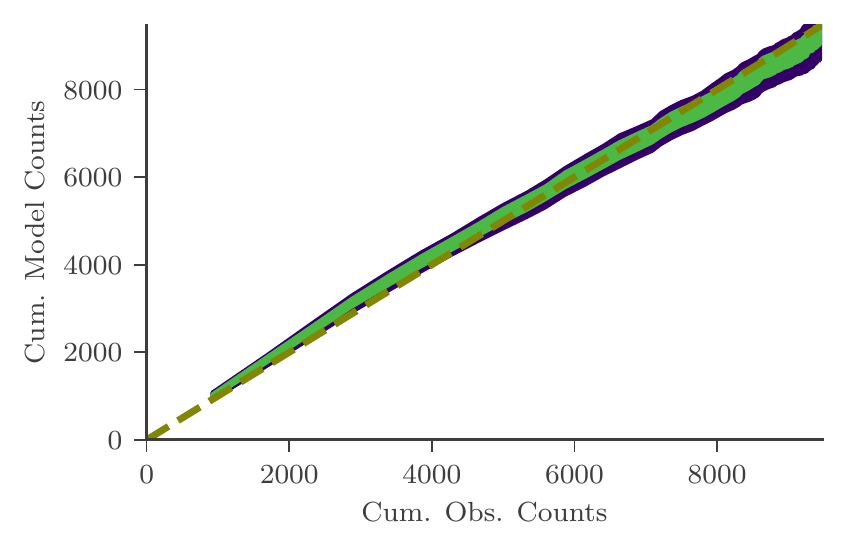}
      \caption{PPC (left) and QQ (right) plot for the GBM detector b0. The green and purple shaded area are the one- and two-sigma contours. The dark yellow curve has the same signification as in Fig. B.1. for the two plots.}
      \label{fig:b0_model_check}
    \end{centering}
  \end{figure*}

  \begin{figure*}[ht]
    \begin{centering}
      \includegraphics[width=0.45\linewidth]{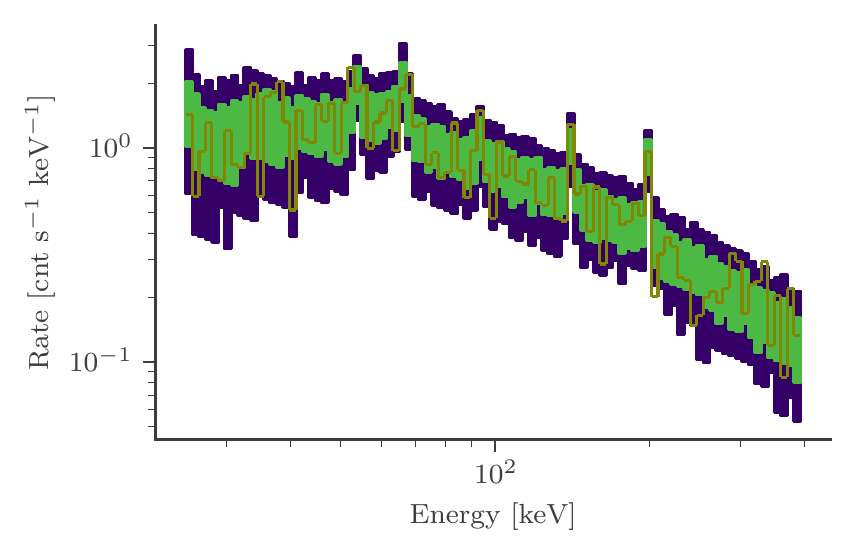}
      \includegraphics[width=0.45\linewidth]{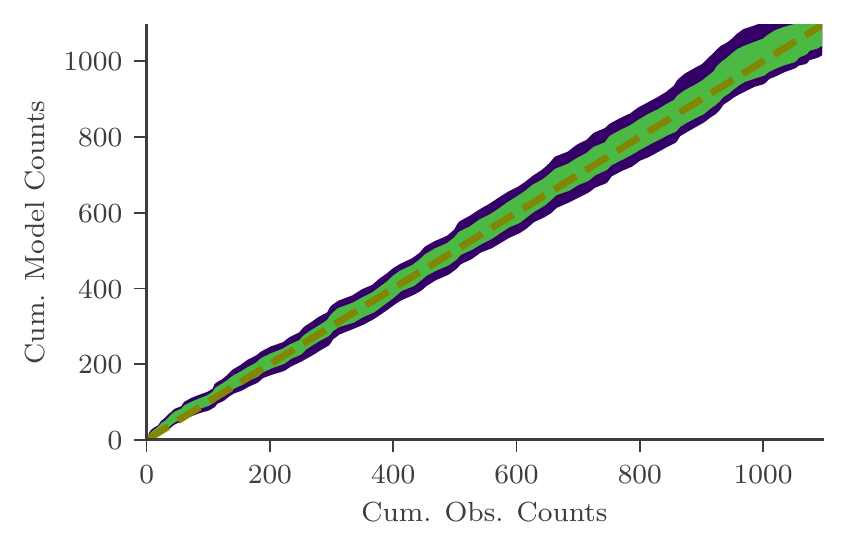}
      \caption{PPC (left) and QQ (right) plot for the low energy range of SPI detector 13 (all single events). The dark
yellow curve has the same signification as in Fig. B.1. for the two plots.}
      \label{fig:low_13_model_check}
    \end{centering}
  \end{figure*}

  \begin{figure*}[ht]
    \begin{centering}
      \includegraphics[width=0.45\linewidth]{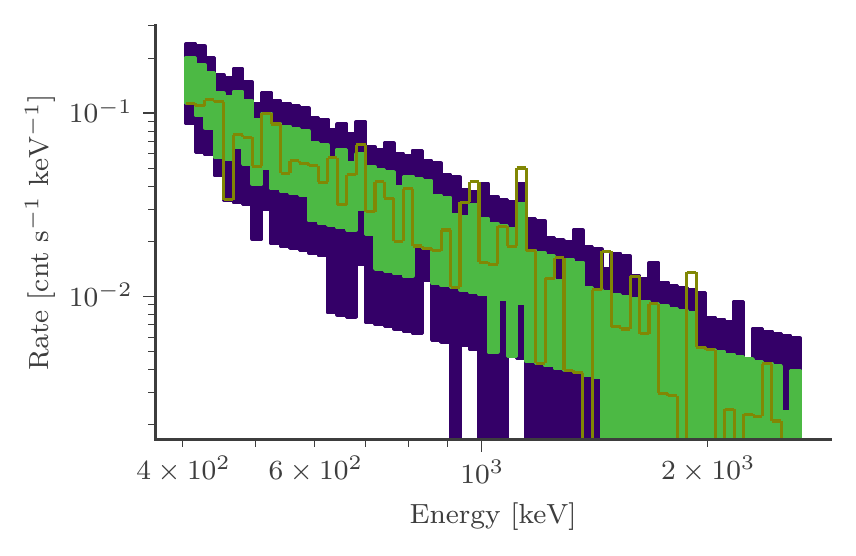}
      \includegraphics[width=0.45\linewidth]{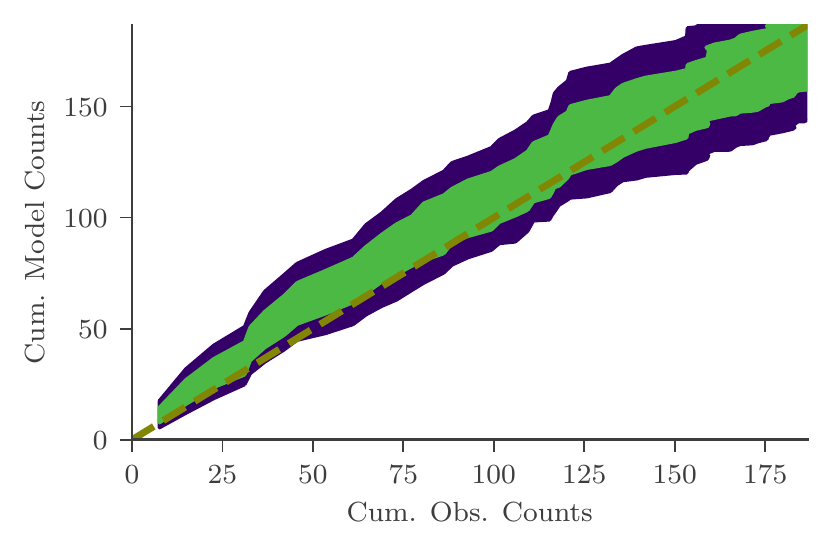}
      \caption{PPC (left) and QQ (right) plot for the middle energy range of SPI detector 13 (only PSD events). The dark
yellow curve has the same signification as in Fig. B.1. for the two plots.}
      \label{fig:psd_13_model_check}
    \end{centering}
  \end{figure*}
\end{appendix}

\end{document}